\documentclass[prl,showpacs,twocolumn,amsmath,amssymb]{revtex4}
\usepackage[dvips]{graphicx}
\usepackage{epsfig}
\usepackage{psfrag}

\begin{document}
\title{Incoherence and enhanced magnetic quantum oscillations in
the mixed state of a layered organic superconductor}
\author{V. M. Gvozdikov$^{1,2}$ and J. Wosnitza$^3$\cite{presadr}}
\affiliation{$^1$Max-Planck-Institut f\"ur Physik komplexer Systeme,
D-01187 Dresden, Germany\\
$^2$Kharkov National University, 61077, Kharkov, Ukraine\\
$^3$Institut f\"ur Festk\"orperphysik, Technische
Universit\"at Dresden, D-01062 Dresden, Germany}

\date{\today}

\begin{abstract}
We present a theory which is able to explain enhanced magnetic
quantum-oscillation amplitudes in the superconducting state of
a layered metal with incoherent electronic transport across the
layers. The incoherence acts through the deformation of the
layer-stacking factor which becomes complex and decreases the
total scattering rate in the mixed state. This novel mechanism 
can compensate the usual decrease of the Dingle factor below the 
upper critical magnetic field caused by the intralayer scattering.
\end{abstract}


\maketitle
In recent years it has been questioned whether the electronic
properties of layered quasi-two-dimensional (2D) metals can be
described within the usual fundamental concept of an anisotropic
three-dimensional Fermi liquid \cite{mck98}. Well-studied examples,
besides the cuprate superconductors and quasi-one-dimensional organic
metals, are organic conductors of the type (BEDT-TTF)$_2X$, where
BEDT-TTF stands for bisethylenedithio-tetra\-thiafulvalene and $X$
for a monovalent anion. This class of materials displays a number
of unique properties such as an unconventional electronic interlayer
transport, clear deviations of their magnetic quantum oscillations
\cite{wos96} from the standard three-dimensional (3D)
Lifshitz--Kosevich theory \cite{lif56}, and puzzling features in the
superconducting mixed state. After the first observation of the de
Haas--van Alphen (dHvA) oscillations in the mixed state of a layered
superconductor \cite{gra76} it was firmly established that these
oscillations are damped below the upper critical field $B_{\rm c2}$.
This damping was explained by several mechanisms reviewed in \cite{man01}.
A recent observation that both the dHvA and the Shubnikov--de Haas
(SdH) amplitudes are enhanced in the mixed state of the layered organic
superconductor $\beta{''}$-(BEDT-TTF)$_2$SF$_5$CH$_2$CF$_2$SO$_3$
($\beta{''}$ salt in the following) \cite{wos03} was a real surprise
since it is in sharp contrast to all experiments and theories known
so far \cite{man01}.

Theoretically, it is assumed that the quasiparticle scattering by the
``vortex matter'' just below the upper critical field, $B_{\rm c2}$,
is the main mechanism of the damping in this region yielding the
Dingle-like additional damping factor $R_{\rm s}=\exp(-2\pi/
\Omega\tau_{\rm s})$. Here, $\Omega$ is the cyclotron frequency and
$\hbar/\tau_{\rm s}$ is the sum of the two terms $\hbar/\tau_{\rm s1}
= \Delta^2(\pi/\mu\hbar\Omega)^{1/2}$, due to the intralayer scattering
at vortices \cite{mak91,ste92,was94}, and $\hbar/\tau_{\rm s2} = \beta
\Delta^2/\mu\Omega\tau_0$, which takes into account the scattering by
impurities and vortices within the layers \cite{gvo98}. ($\tau_{\rm s}$
is the scattering time, $\mu$ is the chemical potential, $\Delta$ is
the superconducting order parameter which just below $B_{\rm c2}$ is
less than the Landau-level separation $\Delta < \hbar\Omega$, and
$\beta \simeq 1$.)

The additional factor $R_{\rm s}$ describes the extra damping of the
dHvA and SdH amplitudes only due to the intralayer scattering in 2D
conductors. Within this concept, however, the oscillation-amplitude
enhancement in the $\beta{''}$ salt is not explicable. Anomalous dHvA
oscillations have also been observed in YNi$_2$B$_2$C \cite{goll96,tera97}.
These oscillations persist down to the surprisingly low field
0.2$B_{\rm c2}$ \cite{tera97}. A Landau-quantization scheme for fields
well below $B_{\rm c2}$ in the periodic vortex-lattice state was developed
in \cite{gork98} and for a model with an exponential decrease of the
pairing matrix elements in \cite{gvo98}. The observed recovery
of the dHvA amplitudes for $B \ll B_{\rm c2}$ in YNi$_2$B$_2$C was
explained by the enhancement of the special vortex-lattice factor
depending on the Landau bands which become narrower when the
vortex-lattice grows thinner \cite{gvo98}.

The anomalous magnetic quantum oscillations in the mixed state of
the $\beta{''}$ salt is much more mysterious and poses the question
on the peculiarity of this material. The $\beta{''}$ salt is the only
among the known superconductors which (i) displays enhanced magnetic
quantum-oscillation amplitudes in the mixed state and (ii) has no 3D
Fermi surface (FS), i.e., exhibits an incoherent electronic transport
across the layers \cite{wos02}. The incoherence means that the electronic
properties of this layered quasi-2D metal cannot be described within the
usual fundamental concept of an anisotropic 3D FS. Nonetheless, magnetic
oscillations due to the 2D FS survive \cite{mck98}.

Here, we consider a new mechanism for the quasiparticle scattering
that goes beyond the usual 2D consideration of {\it intra}layer scattering
by taking into account the {\it inter}layer-hopping contribution to
the total scattering rate. This can explain an oscillation-amplitude
enhancement in the superconducting state for layered conductors with
incoherent hopping across the layers. The clear physical picture behind
this mechanism is as follows. The incoherence, or disorder in the
direction perpendicular to the layers, hampers the electron hopping
between neighboring layers. This enhances the scattering at impurities
within the layers since electrons on a Landau orbit interact with the
same impurities many times before a hopping to the neighboring layer
occurs. In the superconducting state long-range order establishes across
the layers which allows quasiparticles to escape the {\it intra}layer
multiple scattering
by Josephson tunneling between the layers. This mechanism reduces the
scattering rate by impurities and enhances the Dingle factor in the
superconducting state. For the $\beta{''}$ salt this effect most likely
plays the dominant role. For the coherent case, there is no interlayer
scattering and the electrons (quasiparticles) can move freely across the
layers both in the normal and the superconducting state which render the
above mechanism much less effective. Numerically, the effect is described
by the layer-stacking factor [Eq.\ (\ref{3})] which itself contains a
Dingle-like exponent in the case of incoherent interlayer hopping
\cite{gvo84}. Superconductivity restores the coherence across the
layers by renormalizing the hopping integrals \cite{gvo88}. This
reduces the interlayer scattering and enhances the oscillation
amplitudes.

In case the momentum across the layers is not preserved, the electron
interlayer hopping maybe described in terms of an energy $\varepsilon$
that is distributed with the density of states (DOS) $g(\varepsilon)$.
The energy spectrum of a layered conductor in a perpendicular
magnetic field is, therefore, given by $E_n(\varepsilon)=\hbar
\Omega (n+1/2)+\varepsilon$ \cite{gvo84}. The total DOS then
follows from the standard Green-function definition
\begin{equation}
N(E)=\frac{1}{2\pi^{2}l^{2}}\mathit{\rm{Im}}
 \sum_{n=0}^{\infty }\int {\rm d}\varepsilon
 \frac{g(\varepsilon)}{E-E_{n}-\varepsilon-\Sigma_{n}(E)},
\label{1}
\end{equation}
where $\Sigma_n(E)$ is the average self energy corresponding to the
$n$th Landau level and $l=(\hbar e/cB)^{1/2}$ is the magnetic length.
For large energies and large $n\approx E/\hbar\Omega \gg 1$ (relevant
for the magnetic quantum oscillations here) the self energy is
independent on the index $n$ and the summation in Eq.\ (\ref{1})
by use of Poisson's formula yields
\begin{equation}
\frac{N(E)}{N(0)}=1+2{\rm{Re}}
 \sum_{p=1}^\infty (-1)^p R_{\rm D}(p,E)I_p \exp \left(\frac{2\pi ipE}
 {\hbar \Omega }\right),
\label{2}
\end{equation}
where $N(0)$ is the 2D electron-gas DOS and the function $R_{\rm D}(p,E)
=\exp\left(-2\pi p |{\rm{Im}}\Sigma(E)|/\hbar \Omega\right)$
generalizes the Dingle factor to the case of an energy-dependent
self energy $\Sigma(E)$. The layer-stacking factor in Eq.\ (\ref{2}),
\begin{equation}
I_p =\int {\rm d}\varepsilon g(\varepsilon)
\exp \left(\frac{2\pi ip\varepsilon}{\hbar \Omega }\right),
\label{3}
\end{equation}
is an important factor in the theory of magnetic quantum oscillations
in normal and superconducting layered systems \cite{gvo98,gvo02}. It
describes contributions to the oscillations coming from the interlayer
hopping. If the stacking is irregular $I_p$ becomes complex and
contains a Dingle-like exponent \cite{gvo84}. 

The inverse scattering time $1/\tau(E)= |\mathit{\rm{Im}}
\Sigma (E)|/\hbar$ in the self-consistent Born approximation (SCBA) was
found to be proportional to the total DOS \cite{and74,man88,rai93}.
Accordingly, the relation $N(E)/N(0)=\tau_0/\tau(E)$ holds, where
$\tau_0$ is the intralayer scattering time \cite{gri03,cha02}.
Substituting this into Eq.\ (\ref{2}) leads to an equation
for $\tau(E)$ showing that it oscillates as a function of $1/B$.

In case the highly anisotropic electronic system has a 3D Fermi surface
the DOS related to the interlayer hopping is symmetric, $g(\varepsilon)
=g(-\varepsilon)$, and becomes for nearest-neighbor hopping
$g(\varepsilon)=\pi^{-1}(4t^2-\varepsilon^2)^{-1/2}$. The corresponding
layer-stacking factor then is given by $I_p = J_0(4\pi tp/\hbar\Omega)$.
This Bessel function oscillates as a function of $1/B$ which is just
another way to describe the well-known bottle-neck and belly
oscillations of a corrugated 3D FS. Oscillating corrections to
the Ginzburg--Landau expansion coefficients caused by the factor
$J_0(4\pi tp/\hbar \Omega)$ were also calculated in \cite{cha01}.

For the incoherent case, on the other hand, the translation invariance
across the layers is lost. 
The irregular {\it inter}layer hopping means that the DOS deviates from
the function $g(\varepsilon)=\pi^{-1}(4t^2-\varepsilon^2)^{-1/2}$
and loses the symmetry $g(\varepsilon) = g(-\varepsilon)$ 
which implies that ${\rm Im}I_p \ne 0$.
As will be shown below, this results in a special contribution to the
electron scattering time. Below $B_{\rm c2}$, this may lead to a
suppression of the scattering rate acting against the known intralayer
damping mechanisms of the quantum
oscillations in the mixed state \cite{man01}. However, this contribution
vanishes if the hopping between the layers preserves the interlayer
momentum leading to a corrugated 3D FS cylinder. This is an important
point in our consideration.

Using Eq.\ (\ref{2}) and $N(E)/N(0)=\tau_0/\tau(E)$ we write
$\tau(E)^{-1}$ as a sum of the coherent (symmetric) and incoherent
(asymmetric) terms:
\begin{equation}
\tau(E)^{-1}=\tau(E)_{\rm s}^{-1}-\tau(E)_{\rm a}^{-1},
\label{4}
\end{equation}
\begin{equation}
\frac{\tau_0}{\tau_{\rm s}}=1+2
 \sum_{p=1}^\infty (-1)^p R_{\rm D}(p,E){\rm{Re}}I_p
 \cos \left(\frac{2\pi pE}
 {\hbar \Omega }\right),
\label{5}
\end{equation}
\begin{equation}
\frac{\tau_0}{\tau_{\rm a}}=2
 \sum_{p=1}^\infty (-1)^p R_{\rm D}(p,E){\rm{Im}}I_p
 \sin \left(\frac{2\pi pE}
 {\hbar \Omega }\right).
\label{6}
\end{equation}
With the help of the summation rule
\begin{equation}
S(\lambda,\delta)=\sum_{p=-\infty }^{\infty }(-1)^pe^{-|p|\lambda}
\cos p\delta =\frac{\sinh \lambda}{\cosh \lambda +\cos \delta}
\label{7}
\end{equation}
one can rewrite Eqs.\ (\ref{5}) and (\ref{6}) in the integral form
\begin{equation}
\frac{1}{\tau_{\rm s(a)}}=\frac{1}{\tau_{0}}\int {\rm d}\varepsilon
g_{\rm s(a)}(\varepsilon)S[\lambda ,\delta (E,\varepsilon )].
\label{8}
\end{equation}
Here $g_{\rm s}(\varepsilon) = g_{\rm s}(-\varepsilon)$ is the symmetric
and $g_{\rm a}(\varepsilon) = -g_{\rm a}(-\varepsilon)$ is the antisymmetric
part of the DOS $g(\varepsilon)$, $\lambda (E) =2\pi/\Omega \tau$, and
$\delta(E,\varepsilon) =2\pi(E-\varepsilon)/\hbar \Omega$.
The SCBA, as well as Eqs.\ (\ref{4})-(\ref{8}), are valid not only for
point-like impurities but also for a smooth random potential provided
its correlation radius is less than the Larmor radius, which holds for
large $n$ \cite{rai93}.
One can see from Eqs.\ (\ref{4})-(\ref{8}) that, {\it in general}, the
incoherent contribution, $-\tau(E)_{\rm a}^{-1}$, to the total scattering
rate, $\tau(E)^{-1}$, is essential. The integral equation for
$\tau(E)^{-1}$ is very complex and can be solved only perturbatively in
the case $\lambda \gg 1$. In the limit $\lambda \to \infty$, when
$S(\lambda ,\delta) \to 1$, we have $\tau_{\rm s}^{-1}= \tau_0^{-1}$
and $\tau_{\rm a}^{-1}=0$. For finite, but large $\lambda$ the parameter
$R_{\rm D}(p,E)=e^{-p\lambda}\ll 1$. Even if $\Omega\tau \simeq 1$, the
quantity $e^{-\lambda}\ll 1$ and Eqs.\ (\ref{5}) and (\ref{6}) are just
a series expansion in powers of the small parameter $e^{-\lambda}$.
Eq.\ (\ref{7}) shows the convergence of this series for any $\lambda >
0$ allowing a perturbative solution. The perturbative terms oscillate as
a function of $E$ and can be written as the series
$\tau^{-1}(E)=\tau_{0}^{-1}[1+X_1+X_2+ O(e^{-3\lambda_0})]$, with
$X_1\propto e^{-\lambda_0}, X_2\propto e^{-2\lambda_0}$, and
$\lambda_0=2\pi/\Omega \tau_{0}$.
The first nonzero correction averaged over an oscillation period is
proportional to $\overline {X_1+X_2}=-2\lambda_0e^{-2\lambda_0}|I_1|^2$
and yields
\begin{equation}
\frac{1}{\bar{\tau}}=\frac{1}{\tau_0}\left[1-
\frac{4\pi}{\Omega\tau_0}(R^0_{\rm D})^2\left({\rm{Re}}I_1^{2}+
{\rm{Im}}I_1^{2}\right)\right].
\label{9}
\end{equation}
The Dingle factor $R^0_{\rm D} = \exp{(-2\pi/\Omega\tau_0)}$ is a small
parameter in our perturbative solution. The term ${\rm Im}I_1^{2}$ in
Eq.\ (\ref{9}) appears due to the incoherence.

It was established that in the $\beta{''}$ salt electron hopping
across the layers is most probably incoherent, i.e., the momentum
perpendicular to the layers is not preserved and there is no 3D Fermi
surface \cite{wos02}. The reason for this remarkable feature is unknown
so far. It might be that some kind of disorder, such as different
spatial configurations in the extraordinary large and complex
anion-molecule layer may induce random hopping integrals, in analogy
with intercalated layered compounds \cite{gvo84}. Furthermore, the
$\beta{''}$ salt is the only material so far studied (not only among
the BEDT-TTF salts) which displays an enhancement of the magnetic
quantum-oscillation amplitude in the superconducting state \cite{wos03}.

\begin{figure}
\centering
\includegraphics[width=8cm,clip=true]{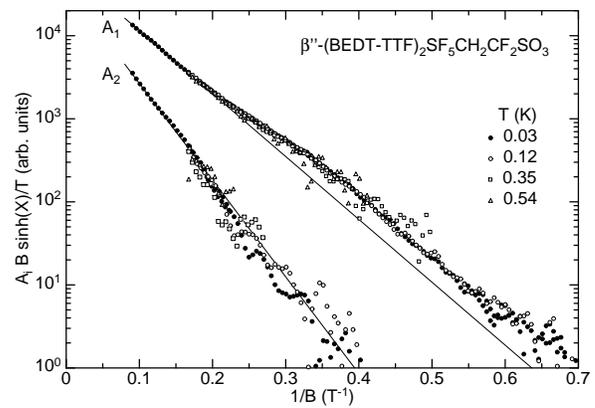}
\caption[]{Dingle plot of the fundamental ($A_1$) and the second harmonic
($A_2$) dHvA amplitudes of the $\beta{''}$ salt extracted from
modulation-field data for different temperatures. The solid lines are fits
to the data above 6~T.} \label{dingle}
\end{figure}

This important result is summarized in Fig.~\ref{dingle}.
In this Dingle plot for a 2D metal the amplitudes $A_1$ of the
fundamental dHvA frequency ($F \approx 198$~T) and $A_2$ of the second
harmonic (2$F$) are normalized by $B\sinh(X)T^{-1}$ and plotted on a
logarithmic scale as a function of $1/B$, with $X = 2\pi^2k_{\rm B}
m_{\rm c}T/e\hbar B$ and $m_{\rm c}$ the effective cyclotron mass
(see \cite{wos03} for more details). Upon entering the superconducting
state the oscillation amplitude $A_1$ is enhanced compared to the
normal-state dependence (solid lines). For the second harmonic $A_2$
neither an additional damping nor an enhancement is observed \cite{a2rem}.

In the superconducting state long-range order across the layers 
evolves through the renormalization of the hopping integrals \cite{gvo88}.
This can be understood as follows. The quasiparticle hopping between the
layers is an independent degree of freedom with respect to the in-plane
Landau quantization. In the normal state the Green-function equation
related to the interlayer hopping is given by
\begin{equation}
\sum_m[(\varepsilon -\varepsilon_i)\delta_{im}-t_{im}]
G^{0}_{ij}(\varepsilon)=\delta_{ij},
\label{10}
\end{equation}
where the electron energy in the layer $\varepsilon_{i}$ and the
hopping integrals $t_{im}=t_{i}(\delta_{i,m+1}+ \delta_{i,m-1})$ are
assumed to depend on the layer indices for the sake of generality.
In the superconducting state the order parameters in the layers,
$\Delta_{i}$, become nonzero and the Gor'kov equation for the
Green functions $G_{ij}$ can be written as \cite{gvo88}
\begin{equation}
\sum_m[(\varepsilon-\varepsilon_i)\delta_{im}-\hat{t}_{im}]
G_{ij}(\varepsilon)=\delta_{ij},
\label{11}
\end{equation}
\begin{equation}
\hat{t}_{im}=t_{im}-\Delta_iG^0_{im}(-\varepsilon)\Delta^\star_m.
\label{12}
\end{equation}
The star means the complex conjugate. When comparing Eq.\ (\ref{10})
with Eq.\ (\ref{11}) it is seen that in the superconducting state
the effective hopping integrals, given by Eq.\ (\ref{12}), become
nonzero not only for next-nearest-neighbor hopping. In that case
the nonvanishing retarded Green-function components $G^0_{im}
(-\varepsilon)$ result in nonzero $\hat{t}_{im}$ for electron hopping
between arbitrary sites $i$ and $m$. In fact, Eq.\ (\ref{12})
simply reflects how the superconducting long-range order establishes
across the layers. The complex order parameters in the layers
$\Delta_i=|\Delta_i|\exp{(i\varphi)}$ appear and result in interlayer
(intrinsic) Josephson coupling \cite{gvo88}. In the absence of
Josephson currents the order parameter can be chosen to be real
and independent of the layer index $\Delta_{i}=|\Delta_{i}|$.
The correction to the DOS due to this mechanism is
\begin{equation}
\delta g(\varepsilon)=-\Delta^{2}\sum_{ij}
\frac{\partial g(\varepsilon)}{\partial t_{ij}}G^0_{ij}(-\varepsilon).
\label{13}
\end{equation}

The second effect we have to take into account is caused by the vortex
matter in the mixed state. Here, the vortices are disordered for fields
slightly below $B_{\rm c2}$. They convert the degenerated Landau levels
into asymmetric Landau bands as was shown in Refs.\ \cite{nor96,tes98}.
Thus, in the mixed state the quasiparticle tunneling between the layers
implies the quantum transition between these Landau-band states which
results in the additional contribution to the simple nearest-neighbor DOS
\begin{equation}
\delta g_\Delta(\varepsilon)=\Delta^{2}\left(
\frac{\partial g(\varepsilon)}{\partial \Delta^2}
\right)_{\Delta=0}.
\label{14}
\end{equation}
The total correction to the DOS in the mixed state can be written as
$\delta g_{\rm tot}(\varepsilon)=\Delta^{2}G(\varepsilon)$, where
$G(\varepsilon)$ is directly defined by Eqs.\ (\ref{13}) and (\ref{14}).
Inserting $\delta g_{\rm tot}(\varepsilon)$ into Eq.\ (\ref{8}) and
averaging over a period in $E$ results in
\begin{equation}
\delta \left(\frac{1}{\tau}\right)=\frac{\Delta^2}{\tau_0}
\int {\rm d}\varepsilon G(\varepsilon)
\overline {S(\lambda,\delta(E,-\varepsilon))}=
\frac{\Delta^2}{\tau_0}\gamma.
\label{15}
\end{equation}
Since the DOS is normalized ($\int {\rm d} \varepsilon g(\varepsilon)=1$)
the function $G(\varepsilon)$ satisfies the condition
$\int {\rm d} \varepsilon G(\varepsilon)=0$. This means that it is
alternating in sign and $\gamma$ might be negative because
$\overline {S(\lambda,\delta(E,-\varepsilon))}>0$.
The studied system is too complex to calculate $\gamma$ in general. 
In the limit $\lambda \to \infty$ this coefficient vanishes since
$\overline {S(\lambda,\delta(E,-\varepsilon))}\to 1$.
It is instructive to consider a correction to the scattering rate in
Eq.\ (\ref{9}) in the mixed state. The variation of the layer-stacking
factor $\delta I_{1}$ is given by Eq.\ (\ref{3}) with the DOS replaced
by $\delta g_{\rm tot}(\varepsilon)$. The broadening of the Landau
levels, caused by $\delta g_{\rm tot}(\varepsilon)$, is of the order
of the width of this function and much less then $\hbar \Omega$ in
order to observe the oscillations. Therefore,
in first approximation ${\rm{Re}}\delta I_{1} \approx 0$ and
${\rm{Im}}\delta I_{1}=\Delta^2 \int {\rm d}\varepsilon G(\varepsilon)
\sin(\frac{2\pi \varepsilon}{\hbar \Omega})\approx
\Delta^2(\frac{2\pi <\varepsilon>}{\hbar \Omega})$, where
$\left<\varepsilon\right>=\int d\varepsilon G(\varepsilon) \varepsilon$.
(For $G(\varepsilon)=-G(-\varepsilon)$, ${\rm{Re}}\delta I_{1}$ is
zero exactly.)
Thus, the correction to the scattering rate in the mixed state near
$B_{\rm c2}$, caused by the interlayer-hopping mechanism, is given by
\begin{equation}
\delta \left(\frac{1}{\bar{\tau}}\right)=
-\Delta^2 \frac{2{\rm{Im}}I_{1}}{\tau_0}
\left(\frac{4\pi}{\Omega\tau_0}\right)(R^0_{\rm D})^2
\left(\frac{2\pi \left<\varepsilon\right>}{\hbar \Omega}\right).
\label{16}
\end{equation}
For $\left<\varepsilon\right>{\rm{Im}}I_1>0$, this gives a decrease
in the scattering rate. Note that the latter is nonzero only if the
system is incoherent in the normal state and ${\rm{Im}}I_1 \ne 0$.
This strongly supports the relevance of this mechanism for the $\beta{''}$
salt, since only this organic metal displays both incoherence in the
normal state and an enhancement of the SdH and dHvA amplitudes in the
mixed state.

Thus, the overall effect superconductivity has on $R_{\rm s}$ is determined
by the balance between positive and negative contributions to the scattering
rate. The already mentioned positive contribution from the intralayer
scattering at vortices and defects is
\begin{equation}
\frac{\hbar}{\tau_{\rm s}}=\Delta^2\left[
\left(\frac{\pi}{\mu\hbar\Omega} \right)^{1/2}+
\frac{\beta}{\mu\Omega\tau_0}\right].
\label{17}
\end{equation}
The new additional interlayer mechanism we discuss
here results in a negative contribution given by $\delta \left(\frac{1}
{\tau}\right) = \frac{\Delta^{2}}{\tau_0}\gamma$ [Eq.(\ref{15})]. Since
little is known about the DOS of the studied system, even in the normal
state, we cannot calculate the coefficient $\gamma$ quantitatively.
However, contrary to Eq.\ (\ref{17}), for this term
the small factor $1/\mu$ is absent, so that the overall correction
to the scattering rate might be negative.
The experimental facts \cite{wos03} give us confidence that this
is the case at least for the $\beta{''}$ salt.

We conclude with a qualitative picture of the effect discussed here.
The incoherence means that the hopping time between the layers
$\tau_z\approx\hbar/|t| \gg \tau_0$ so that an electron scatters many
times within a layer before leaving it \cite{mck98}. Here the quantity
$|t|$ is some averaged hopping integral that in the $\beta{''}$ salt
may be assumed to be the smallest parameter in energy. Indeed,
experimentally $|t|$ cannot be resolved in the $\beta{''}$ salt
reflecting the fact that the hopping integral is one of the smallest
for all known 2D organic metals so far \cite{wos02}. Consequently, even
small spatial fluctuations of the hopping probability within and across
the layers render the electron motion across the layers incoherent.
On the other hand, for the evolution of superconductivity some
interlayer (Josephson) coupling is vitally important. Long-range
order establishes below $B_{\rm c2}$, thereby renormalizing the hopping
integral. According to Eq.~(\ref{12}), the renormalized $\tau_z$ in the
superconducting state may be estimated as $\tau_z\approx\hbar/|t+
\Delta^{2}/t|$. For $\Delta \gg|t|$ the hopping time reduces
considerably and becomes $\tau_z \approx \hbar |t|/\Delta^2$. The
latter means that the quasiparticles spend less time within the
(impurity-containing) layers decreasing the scattering rate and,
consequently, enhancing the Dingle factor. In the $\beta{''}$ salt
this effect is strong because of the smallness of $|t|$. Thus, our
mechanism relates the two unusual effects observed in the $\beta{''}$
salt: the incoherent interlayer hopping transport and the enhancement
in the quantum-oscillation amplitudes in the mixed state.

\acknowledgments
This work was supported in part by INTAS, project INTAS-01-0791, and
the NATO Collaborative Linkage Grant No.\ 977292.
V.M.G.\ thanks P.\ Fulde and S.\ Flach for the hospitality at the 
MPIPKS in Dresden.


\begin{thebibliography}{123}
\bibitem{a2rem}
A different behavior of $A_2$ in the mixed state is not surprising
since for 2D metals, and especially for the $\beta''$ salt,
$A_2$ behaves unusual even in the normal state.
\end{thebibliography}
\end{document}